\documentclass[aps,showpacs,10pt]{revtex4}

\usepackage{graphicx}
\usepackage{amsmath}
\usepackage{amssymb}

\newcommand{\be}{\begin{equation}}
\newcommand{\ee}{\end{equation}}
\newcommand{\ba}{\begin{eqnarray}}
\newcommand{\ea}{\end{eqnarray}}

\def\bk{{\bf k}}

\newcount\bozza \bozza=0
\ifnum\bozza=1
\newdimen\shift \shift=-2truecm
\def\lb#1{%
{\label{#1}\rlap{\kern\shift{$\scriptstyle#1$}}}}
\else\def\lb#1{\label{#1}} \fi

\begin{document}


\title{Superfluid density and competing orders in $d$-wave superconductors}

\author{S.G.~Sharapov}
\email{sharapo@mcmaster.ca}
\author{J.P.~Carbotte}
\email{carbotte@mcmaster.ca}

\affiliation{Department of Physics and Astronomy, McMaster University,
        Hamilton, Ontario, Canada, L8S 4M1}

\date{\today }

\begin{abstract}
We derive expressions for the superfluid density $\rho_s$ in the
low-temperature limit $T \to 0$ in $d$-wave superconductors, taking
into account the presence of competing orders such as spin-density
waves, $i d_{xy}$-pairing, etc. Recent experimental data for the
thermal conductivity and for elastic neutron scattering in
La$_{2-x}$Sr$_x$CuO$_4$ suggest there are magnetic field induced
anomalies that can be interpreted in terms of competing orders. We
consider the implications of these results for the superfluid
density and show in the case of competing spin-density wave order
that the usual Volovik-like $\sqrt{H}$ depletion of $\rho_s(H)$ is
replaced by a slower dependence  on applied magnetic field. We find
that it is crucial to include the competing order parameter in the
self-consistent equation for the impurity scattering rate.
\end{abstract}




\maketitle

\section{Introduction}

The doping and temperature dependence of the superfluid density
$\rho_s(x,T)$ and its correlation with the critical temperature
$T_c$ in high-temperature superconductors (HTSC) have been
intensively studied since their discovery \cite{Uemura1991PRL}.
The last few years  are also marked by new observations
\cite{Tallon2003PRB,Homes2004Nature,Zaanen2004Nature,Homes2004}
particularly for the underdoped side of the phase diagram
\cite{Liang2005PRL,Zuev2005PRL,Micnas2002JPCM,Herbut2005PRL}. This
special attention payed to the superfluid density is not
surprising, because all  properties of $\rho_s(x,T)$ mentioned
above show that HTSC differ significantly from previously known
superconductors. An important example is the temperature
dependence of the in-plane penetration depth $\lambda(T)$
\cite{foot1} which provided one of the first strong evidence
\cite{Hardy1993PRL} for unconventional pairing symmetry and for
the important role of gapless low-energy quasiparticle
excitations.

Another interesting dependence of $\rho_s$ is its variation with the
external magnetic field $H$. This dependence contains information on
the quasiparticles and on their interaction with the vortices
present in the vortex phase of HTSC. Since these are extreme type-II
superconductors, a huge vortex phase extends from the lower critical
field, $H_{c1} \sim 10 - 100 \, \mbox{Gauss}$ to the upper critical
field, $H_{c2} \sim 100 \, \mbox{T}$. The leading term in the field
dependence of $\rho_s(H) \sim \sqrt{H}$ is well understood
Refs.~\cite{Vekhter1999PRB,Sharapov2002PRB,Ioffe2002JPCS} using the
local Doppler-shift approximation. This very useful approximation
was introduced by Volovik \cite{Volovik1993JETPL}, who predicted,
that in contrast to conventional superconductors, in $d$-wave
systems the density of states (DOS) is dominated by the
contributions from the excited quasiparticle states rather than the
bound states associated with vortex cores. It was shown that in an
applied magnetic field $H$ the extended quasiparticles DOS,
$N(\omega =0, H) \sim \sqrt{H}$ rather than $\sim H$ of the
conventional case. The validity of the Doppler-shift approximation
was discussed in Refs.~\cite{Dahm2002PRB,Laiho2004PRB} using the
quasiclassical Eilenberger equations. It was shown
Ref.~\cite{Laiho2004PRB} that for the superfluid density this
approximation works reasonably well at low temperatures.

The characteristic $\sqrt{H}$ behavior has been observed in the
specific heat and thermal conductivity (see
Ref.~\cite{Hussey2002AP} for a review of experiments). The
dependence $\rho_s(H)$ can also be directly extracted from
mutual-inductance technique measurements
\cite{Jeanneret1989APL,Zuev2005PRL,Martinoli.report} or, for
example, from muon spin rotation ($\mu$SR) measurements (see
Ref.~\cite{Sonier2000RMP} for a review). A big advantage of the
former method is that it provides directly the desired dependence
of $\rho_s(H)$. Moreover, the measured inductance is directly
related to the superfluid density, so that in contrast to the
thermal conductivity  discussed below there is no need to subtract
a phonon contribution from the raw data \cite{Hussey2002AP}. On
the other hand, the analysis of $\mu$SR experiments involves
additional assumptions about the internal magnetic field
distribution in the vortex state \cite{Sonier2000RMP} which are
contained in a model dependent function $f(H)$.  This important
field dependent function enters the relationship $\sigma(H) = f(H)
\rho_s(H)$ between the measured  muon depolarization rate
$\sigma(H)$ and $\rho_s(H)$.

Since 1997 there is ongoing discussion (see
Ref.~\cite{Varelogiannis2005PRB} for a historical overview) about
the dependence of the thermal conductivity $\kappa(H)$ and its
deviations from the expected $\sim \sqrt{H}$ behavior. The crucial
fact is that when the magnetic field is applied perpendicular to
the CuO$_2$ planes, the thermal conductivity shows a transition
from a field-dependent regime  $\kappa(H) \sim \sqrt{H}$ to a
field-independent, plateau-like regime. The latest experimental
results in underdoped La$_{2-x}$Sr$_x$CuO$_4$ (LSCO)
\cite{Takeya2002PRL,Sutherland2003PRB} at $H=0$ for $0 \leq x \leq
0.22$ and \cite{Sun2003PRL,Hawthorn2003PRL} at $0< H <16 \mbox{T}$
for $0.06 \leq x \leq 0.22$ are interpreted in
Ref.~\cite{Hawthorn2005} in terms of a {\em competing
spin-density-wave order\/} in the underdoped regime.

A theoretical background for this interpretation was proposed in
Ref.~\cite{Gusynin2004EPJB}, where it was shown that in the presence
of a spin-density wave gap $m$ \cite{foot2} for $H=0$ at $T\to0$
\begin{equation}
\label{thermal-VP} \frac{\kappa}{T} = \frac{k_B^2}{3} \frac{v_F^2
+v_\Delta^2}{v_F v_\Delta} \frac{\Gamma_0^2}{\Gamma_0^2 + m^2}.
\end{equation}
Here $v_F$ is the Fermi velocity, $v_\Delta$ is the gap velocity and
$\Gamma_0$ is the impurity scattering rate which in
Ref.~\cite{Gusynin2004EPJB} is assumed to be independent of the gap
$m$ (we use units with $\hbar = c=1$, unless stated explicitly
otherwise, and from Sec.~\ref{sec:formalism} also set $k_B =1$).

As one can see from Eq.~(\ref{thermal-VP}) for $m=0$ and $v_F =
v_{\Delta}$ the minimal value of the thermal conductivity is
$\kappa_{min}/T = 2 k_B^2/3$. The opening of the gap $m$ leads to a
suppression of the thermal conductivity and allows for values of
$\kappa/T$ which are less than $\kappa_{min}/T$, as is indeed
observed in LSCO for $x=0.06$ \cite{Sutherland2003PRB}. The presence
of a nonzero field-dependent gap $m(H)$ also allows one to explain
the behavior of $\kappa(H,m(H))$ observed in underdoped LSCO
\cite{Sun2003PRL,Hawthorn2003PRL}.

There are two additional experiments that support the idea of the
presence of a competing order in LSCO. The first is an
angle-resolved photoemission study \cite{Shen2004PRB} that indicates
the existence of a finite gap over the entire Brillouin zone,
including the $d_{x^2-y^2}$ nodal line in LSCO for $x<0.03$. And an
even more exciting observation was made for  a LSCO sample with
$x=0.144$ and $T_c = 37 \mbox{K}$ by elastic neutron scattering
\cite{Khaykovich2005PRB} which showed that a static incommensurate
spin-density-wave order develops above a critical field $H_{0}
\approx 3 \mbox{T}$. This picture is supported by the latest
measurements \cite{Li2005} made on a sample with the same doping
which showed  $\kappa(H)/T$ increasing for $0<H\lesssim 0.5
\mbox{T}$ and decreasing at higher fields. However, a more heavily
doped LSCO sample with $x=0.15$ does not show any decrease of
$\kappa(H)/T$ up to 17T. All these observations fuel interest in the
investigation of competing orders.

The purpose of the present work is to investigate the influence of
competing orders on the superfluid density $\rho_s$ and its field
dependence. We demonstrate that it is crucial to include the effect
of the opening of the spin-density-wave gap $m$ on the value of the
impurity scattering rate $\Gamma(m)$. When this effect is taken into
account the the resulting dependence $\rho_s(H, m(H),\Gamma(m(H)))$
resembles the experimental results on very thin films of LSCO
\cite{Martinoli.report}.

Although the latest experiments on LSCO
\cite{Khaykovich2005PRB,Li2005} are mainly interpreted
\cite{Hawthorn2005} in terms of a competing spin density wave
(SDW) order, the earlier experiments on BSCO (see Refs. in
\cite{Varelogiannis2005PRB} and Ref.~\cite{Movshovich1998PRL})
were also interpreted using a {\em complex\/} $d_{xy}$ component
generated in a $d$-wave superconductor in the magnetic field
\cite{Laughlin1998PRL,Balatsky2000PRB}. The removal of the
$d$-wave node  in optimally doped YBCO in a magnetic field was
observed in the in-plane tunneling conductance \cite{Beck2004PRB}.
The authors of Ref.~\cite{Beck2004PRB} interpreted their
observation in terms of competing $i d_{xy}$ or $i s$ order. Here
we also consider the influence of competing superconducting $i
d_{xy}$ and $i s$ orders on the superfluid density $\rho_s$. This
problem was considered previously by Modre {\em et al.}
\cite{Modre1998PRB} who calculated the London penetration depth
for mixed superconducting order in zero magnetic field.  We point
out that the superfluid density $\rho_s$ may provide a way to
distinguish SDW and superconducting orders.

The paper is organized as follows. In Sec.~\ref{sec:formalism} we
introduce the $4\times4$ Dirac formalism convenient for the
description of competing order with the underlying $d$-wave
superconductivity. The general representation for the superfluid
density is written in Sec.~\ref{sec:superfluid} and in
Sec.~\ref{sec:rho_s-formal} the difference between $\rho_s(T=0)$ for
competing (with $d$-wave superconductivity) superconducting and SDW
orders is considered. The dependence of the impurity scattering rate
$\Gamma$ on the values of competing gaps both in the Born and
unitary limits is discussed in Sec.~\ref{sec:impurity}. In
Sec.~\ref{sec:superfluid_gap} we derive analytical expressions for
$\rho_s(T)$ in the presence of competing orders at $H=0$ in the
low-temperature limit. The main results of the paper for the field
dependence $\rho_s(H)$ at $T=0$ are presented in
Sec.~\ref{sec:superfluid-vortex} In Conclusions,
Sec.~\ref{sec:concl} we give a summary of the results obtained.

\section{Dirac formalism for description of competing orders in $d$-wave
superconductors} \label{sec:formalism}

We begin with the action for a $d$-wave superconductor written in
imaginary time - momentum representation
\begin{equation}\label{action-Nambu}
S = -\int d \tau \int d \mathbf{k} \psi^\dagger(\tau,\mathbf{k})
\left[\hat{I} \partial_\tau + \tau_3 \xi(\mathbf{k}) - \tau_1
\Delta(\mathbf{k}) \right] \psi(\tau,\mathbf{k}),
\end{equation}
where
\begin{equation}
\Psi^\dagger(\tau,\mathbf{k}) = (c_{\uparrow}^\dagger(\tau,
\mathbf{k}), c_{\downarrow}(\tau,-\mathbf{k}) )
\end{equation}
is the Nambu spinor and $c_{\sigma}^\dagger(\tau, \mathbf{k})$ and
$c_{\sigma}(\tau, \mathbf{k})$ with $\sigma = \uparrow, \downarrow$
are, respectively, creation and annihilation operators. Most of the
time we will rely on the nodal approximation, so that the precise
form of the dispersion law of the quasiparticles $\xi(\mathbf{k})$
and the $d$-wave superconducting gap $\Delta(\mathbf{k})$ is not
essential.

\begin{figure}[h]
\centering{
\includegraphics[width=8cm]{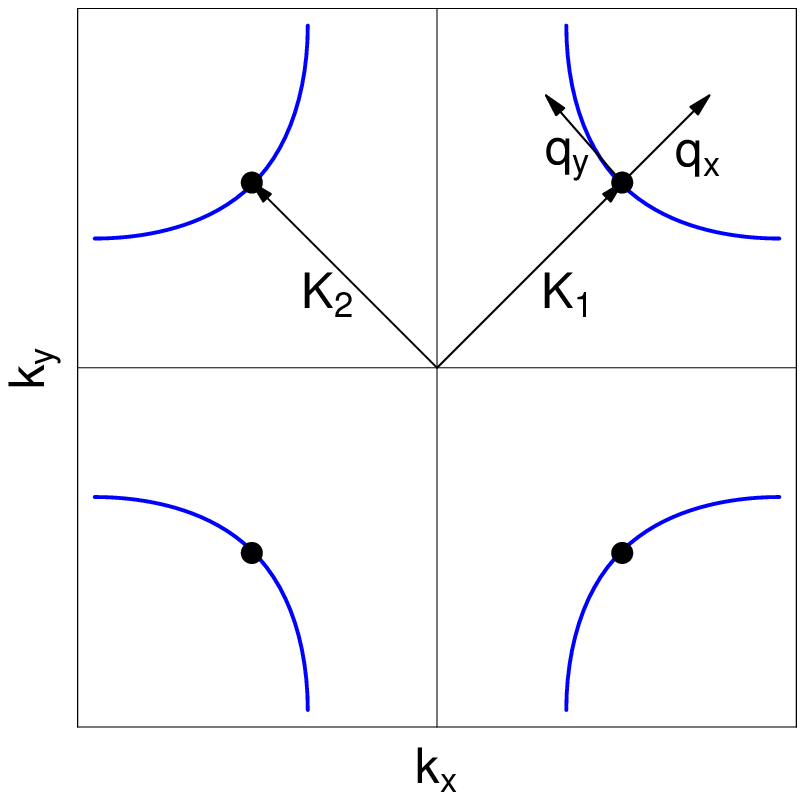}}
\caption{(Color online) The schematic graph of the Fermi surface
with the vectors $\mathbf{K}_i$, $i=1,2$, and $\mathbf{q}$. The SDW
ordering vectors are $\mathbf{Q}_i =2 \mathbf{K}_i$.} \label{fig:1}
\end{figure}

It is impossible to consider other than $is$ competing order while
remaining within a $2\times 2$ formalism. Depending on the physical
assumptions made about the nature of the competing order , there are
different possibilities for constructing a four-component field from
Nambu spinors and switching to a $4\times4$ formalism (see
e.g.~\cite{Vojta2000PRL,Khveshchenko2001PRL,Tesanovic2002PRB,Herbut2002PRB}).
Since we are mostly interested in competing spin density wave which
forms on top the superconducting state, we choose our spinors as was
done in Refs.~\cite{Herbut2002PRB,Gusynin2004EPJB}
\begin{equation}
\label{New-Nambu.variables} \Psi_i^{\dagger}(t, \mathbf{k}) = \left(
\begin{array}{cccc} c_{\uparrow}^{\dagger}(t, \mathbf{k}) \quad
c_{\downarrow}(t, - \mathbf{k}) \quad c_{\uparrow}^{\dagger}(t,
\mathbf{k} - \mathbf{Q}_i) \quad c_{\downarrow}(t, - \mathbf{k} +
\mathbf{Q}_i)
\end{array} \right),
\end{equation}
where $\mathbf{Q}_i = 2 \mathbf{K}_i$ is the wave vector that
connects the nodes within the diagonal pair $i=1,2$. Further since
we are interested in the low-temperature ($T \ll T_c$) properties of
the system, we consider only the vicinity of the nodes $\mathbf{k}
=\mathbf{K}_i +\mathbf{q}$ with $|\mathbf{q}| \ll |\mathbf{K}_i|$ as
shown in Fig.~\ref{fig:1}. Using that $\xi(\mathbf{k}) =
-\xi(\mathbf{k}-\mathbf{Q}_i)$, and $\Delta(\mathbf{k}) =
-\Delta(\mathbf{k}-\mathbf{Q}_i)$ for $\mathbf{k} \approx
\mathbf{K}_i$, and then linearizing the spectrum as $\xi(\mathbf{k})
= v_F q_x + O(q^2)$ and $\Delta(\mathbf{k}) = v_\Delta q_y +
O(q^2)$, one arrives at the low-energy action \cite{Herbut2002PRB}
\begin{equation}
\label{action-matrix} S  = -\int d \tau \int d \mathbf{k}
\Psi_1^{\dagger}(\tau, \mathbf{k}) \left[\hat{I}_4
\partial_\tau  + M_1 v_F q_x + M_2 v_\Delta q_y    \right]
\Psi_1(\tau, \mathbf{k}) + (1\longrightarrow2, x\longleftrightarrow
y),
\end{equation}
where $M_1 = \sigma_3 \otimes \tau_3$ and $M_2 = -  \sigma_3 \otimes
\tau_1$, respectively. It is useful to reformulate the model
(\ref{action-matrix}) in the form of QED$_{2+1}$, because this
allows us to rely on the algebraic properties of $4\times4$
reducible representation  of $\gamma$-matrices which satisfy
Clifford (Dirac) algebra
\begin{equation}
\label{algebra} \{\gamma_\mu,\gamma_\nu\} = 2 \hat{I}_4 g_{\mu \nu},
\qquad g_{\mu \nu} = \mbox{diag}(1,-1,-1), \quad \mu,\nu =0,1,2.
\end{equation}
Another advantage of the QED$_{2+1}$ formulation is that it also
allows us to classify different competing orders in terms of
different types of Dirac masses
\cite{Vojta2000PRL,Khveshchenko2001PRL,Tesanovic2002PRB,Herbut2002PRB,Gusynin2004EPJB}.
We introduce the Dirac conjugated spinor $\bar \Psi_i =
\Psi_i^\dagger \gamma_0$, where $\gamma_0$ is the $4 \times4$ matrix
that anticommutes with $M_1$ and $M_2$ and  such that $\gamma_0^2 =
\hat{I}_4$. This choice is not unique, but we will follow the same
conventions as in Refs.~\cite{Herbut2002PRB,Gusynin2004EPJB}
choosing $\gamma_0= \sigma_1 \otimes \tau_0$. Accordingly we define
$\gamma_{1,2}$ via $M_1 = \gamma_0 \gamma_1$ and $M_2 = \gamma_0
\gamma_2$, so that $\gamma_1 = -i \sigma_2 \otimes \tau_3$ and
$\gamma_2 = i \sigma_2 \otimes \tau_1$ satisfy Eq.~(\ref{algebra}).
Finally we arrive at the action
\begin{equation} \label{Dirac-action} S  = - \int
d \tau \int d \mathbf{k} \bar \Psi_1(\tau, \mathbf{k})
\left[\gamma_0
\partial_\tau  + \gamma_1 v_F q_x + \gamma_2 v_\Delta q_y    \right]
\Psi_1(\tau, \mathbf{k}) + (1\longrightarrow2,x\longleftrightarrow
y).
\end{equation}

One may consider quasiparticle gaps $m_i$ of different nature
encoding it in the matrix structure $O_i =(\hat{I}_4, i \gamma_5,
\gamma_3, \gamma_3\gamma_5)$. Here the matrices $\gamma_3$ and
$\gamma_5$, anticommuting with the matrices $\gamma_\nu$, and are
\begin{equation}
\gamma_3 = i \sigma_2 \otimes \tau_2, \qquad \gamma_5 = \sigma_3
\otimes \hat{I}_2.
\end{equation}
Then, different gaps $m_i$ correspond to different types of Dirac
masses added to the action (\ref{Dirac-action}). In particular, the
mass $m_1$, with $O_1 = \hat{I}_4$, describes the (incommensurate)
$\cos$ spin-density-wave (SDW), and the mass $m_2$ with $O_2 = i
\gamma_5$, describes $\sin$ SDW. The masses $m_3$ and $m_4$ with
$O_3 = \gamma_3$ and $O_4 = \gamma_3 \gamma_5$, correspond to the $i
d_{xy}$-pairing and the $i s$-pairing, respectively
\cite{Herbut2002PRB,Gusynin2004EPJB}. In the present paper we
concentrate mainly on the SDW gap, $m_1 \equiv m$ with the
corresponding bare Matsubara Green's function
\begin{equation}
\label{GF_I} G_0(i\omega_n, \mathbf{k})  = - \frac{i\omega_n
\gamma_0 - v_F k_1 \gamma_1 - v_{\Delta}k_2 \gamma_2+ m
\hat{I}}{\omega_n^2 + v_F^2 k_1^2 + v_{\Delta}^2 k_2^2 + m^2}
\end{equation}
and on $i d_{xy}$ gap, $m_3 = \Delta_{d_{xy}}$ with the
corresponding Green's function
\begin{equation}
\label{GF_id} G_0(i\omega_n, \mathbf{k})  = - \frac{i\omega_n
\gamma_0 - v_F k_1 \gamma_1 - v_{\Delta}k_2 \gamma_2-
\Delta_{d_{xy}} \gamma_3}{\omega_n^2 + v_F^2 k_1^2 + v_{\Delta}^2
k_2^2 + \Delta_{d_{xy}}^2}.
\end{equation}

Finally we should define the electric current operator in $4 \times
4$ formalism. In Nambu formalism it reads
\begin{equation}
\mathbf{j}(\tau, \mathbf{q}=0) = e \int d \mathbf{k}
\mathbf{v}_F(\mathbf{k}) \psi^\dagger(\tau,\mathbf{k}) \hat{I}_2
\psi(\tau, \mathbf{k}), \qquad \mathbf{v}_F(\mathbf{k}) \equiv
\frac{\partial \xi(\mathbf{k})}{\partial \mathbf{k}}, \qquad d \bk
\equiv \frac{d^2 k}{(2 \pi)^2},
\end{equation}
and $\xi(\mathbf{k}) = -\xi(\mathbf{k}-\mathbf{Q}_i)$, we arrive at
the expression
\begin{equation}
\begin{split}
\mathbf{j}(\tau, \mathbf{q}=0) & = e \int_{\mathrm{HBZ}} d
\mathbf{k} \mathbf{v}_F(\mathbf{k}) \Psi^\dagger(\tau,\mathbf{k})
\sigma_3
\otimes \hat{I}_2 \Psi(\tau, \mathbf{k}) \\
&= e \int_{\mathrm{HBZ}} d \mathbf{k} \mathbf{v}_F(\mathbf{k}) \bar
\Psi(\tau,\mathbf{k}) \gamma_0 \gamma_5 \Psi(\tau, \mathbf{k}),
\end{split}
\end{equation}
where the integration is over the halved Brillouin zone (HBZ),
i.e., over the domain with $k_y>0$ in Fig.~\ref{fig:1}. This
implies that after the nodal approximation is used, one should
include in the integration only two neighboring nodes with $i=1,2$
as reflected in Eq.~(\ref{Dirac-action}), because the opposite
nodes are already included in the $4\times4$ formalism.

\section{General representation for the superfluid density}
\label{sec:superfluid}

The superfluid stiffness (or density divided by the carrier mass
$m^\ast$) is given by \cite{Durst2000PRB}
\begin{equation}\label{stiffness.def}
\Lambda_{s}^{ij}(T,H) \equiv \frac{\rho_s^{ij}}{m^\ast}  = \tau_{ij}
- \Lambda_n^{ij}(T,H),
\end{equation}
where $\tau_{ij}$ is the diamagnetic (or stress) tensor and
$\Lambda_s$ is related to the London penetration depth $\lambda_L$
by the standard expression $\Lambda_s = c^2/(4 \pi e^2
\lambda_L^2)$. In Eq.~(\ref{stiffness.def}) $\Lambda_n$ is the
normal fluid density divided by the carrier mass, calculated within
the ``bubble approximation'' with dressed fermion propagators (i.e.,
with self-energy $\Sigma$ due to the scattering on impurities
included) but neglecting  vertex and Fermi liquid corrections,
\begin{equation}\label{rho_n}
\Lambda_n^{ij} = - T \sum_{n=-\infty}^{\infty} \int_{\mathrm{HBZ}}
\frac{d^2 k}{(2 \pi)^2} v_{Fi}(\mathbf{k})
v_{Fj}(\mathbf{k})\mbox{tr}[G(i \omega_n,\mathbf{k})\gamma_0
\gamma_5 G(i \omega_n,\mathbf{k}) \gamma_0 \gamma_5].
\end{equation}
As shown in  \cite{Durst2000PRB} the vertex corrections can be
neglected if the impurity scattering potential is isotropic in
$\mathbf{k}$-space. Likewise the Fermi liquid corrections can be
taken into account along the lines of Ref.~\cite{Durst2000PRB}. When
$\Sigma=0$ it is more convenient to begin with the sum over
Matsubara frequencies in Eq.~(\ref{rho_n}) as done in
Sec.~\ref{sec:rho_s-formal} below. In a more generic case $\Sigma
\neq 0$, the momentum integration has to be done first. Using the
spectral representation for the Green's function $G(i \omega_n,
\bk)$ with the spectral function given by the discontinuity of the
fermion Green's function
\begin{equation}
\label{disc} A(\omega,\mathbf{k}) = - \frac{1}{2\pi
i}\left[G_{R}(\omega+i0, \mathbf{k}) -G_{A}(\omega- i
0,\mathbf{k})\right],
\end{equation}
one can easily sum over Matsubara frequencies in Eq.~(\ref{rho_n})
and represent $\Lambda_n$ in the form \cite{Sharapov2002PRB}
\begin{equation}\label{rho.n-RA}
\begin{split}
& \Lambda_{n}^{i j} = \int_{\mathrm{HBZ}} \frac{d^2 k}{(2 \pi)^2}
\int_{-\infty}^{\infty} d \omega  \tanh \frac{\omega}{2 T}
\frac{ v_{F i} v_{F j}}{4 \pi i} \\
& \times \mbox{tr}[ G_{A}(\omega, \mathbf{k})\gamma_0 \gamma_5
G_{A}(\omega,\mathbf{k})\gamma_0 \gamma_5 - G_{R}(\omega,
\mathbf{k}) \gamma_0 \gamma_5G_{R}(\omega, \mathbf{k})\gamma_0
\gamma_5]
\end{split}
\end{equation}
which is more convenient for the integration over $\bk$. In
Eqs.~(\ref{disc}) and (\ref{rho.n-RA}) $G_{R,A}(\omega,\bk) =
G(i\omega_n \to \omega \pm i0,\bk)$ are retarded and advanced
Green's functions.

\subsection{Properties of $\Lambda_s$ for different competing orders}
\label{sec:rho_s-formal}

To simplify the formal consideration, in this section we will assume
that there is  perfect nesting, $\xi(\mathbf{k}) =
-\xi(\mathbf{k}-\mathbf{Q}_i)$ and $\Delta(\mathbf{k}) =
-\Delta(\mathbf{k}-\mathbf{Q}_i)$ for all values of the momentum
$\bk$ and not only in the vicinity of nodes. Then for the
diamagnetic tensor we obtain
\begin{equation}\label{tau}
\begin{split}
\tau_{ij} &= T \sum_{n=-\infty}^{\infty} \int_{\mathrm{HBZ}}
\frac{d^2 k}{(2 \pi)^2} \mbox{tr} [\sigma_3 \otimes \tau_3
G(i\omega_n, \mathbf{k})\gamma_0]\frac{\partial^2
\xi(\mathbf{k})}{\partial k_i
\partial k_j}\\ & = - \int_{\mathrm{HBZ}} \frac{d^2 k}{(2
\pi)^2} \frac{\partial^2 \xi(\mathbf{k})}{\partial k_i
\partial k_j} \frac{2 \xi(\mathbf{k})}{E(\mathbf{k})} \tanh
\frac{E(\mathbf{k})}{2T},
\end{split}
\end{equation}
where depending on the kind of competing order $E(\mathbf{k})=
\sqrt{\xi^2(\mathbf{k})+ \Delta^2(\mathbf{k}) + m^2 +
\{\Delta_{d_{xy}}^2\}}$. Here and in what follows the notation $m^2+
\{\Delta_{d_{xy}}^2\}$ implies that either the competing SDW order
with the gap $m$ or the competing $id_{xy}$ order with the gap
$\Delta_{d_{xy}}$ is considered.

For the second term of Eq.~(\ref{stiffness.def}) evaluating the
trace in Eq.~(\ref{rho_n}) we obtain
\begin{equation}\label{rho_n-clean}
\begin{split}
\Lambda_n^{ij} = - T \sum_{n=-\infty}^{\infty} \int_{\mathrm{HBZ}}
\frac{d^2 k}{(2 \pi)^2} v_{Fi}(\mathbf{k}) v_{Fj}(\mathbf{k})
\frac{4[-\omega_n^2 - m^2 + \{\Delta^2_{d_{xy}}\} +
\xi^2(\mathbf{k})+ \Delta^2(\mathbf{k})]}{[\omega_n^2
+\xi^2(\mathbf{k})+ \Delta^2(\mathbf{k})+m^2 + \{
\Delta^2_{d_{xy}}\}]^2}.
\end{split}
\end{equation}
One may notice that there is an important difference between SDW and
$id_{xy}$ cases in the sign before $m^2$ and $\Delta_{d_{xy}}^2$ in
the numerator of the Matsubara sum. Evaluating the sum firstly for
$\Delta_{d_{xy}}^2$ we arrive at
\begin{equation}\label{rho_n-SC}
\Lambda_n^{ij} = \int_{\mathrm{HBZ}} \frac{d^2 k}{(2 \pi)^2}
v_{Fi}(\mathbf{k}) v_{Fj}(\mathbf{k}) \frac{1}{T} \frac{1}{\cosh^2
\frac{E(\mathbf{k})}{2T}}, \qquad E(\mathbf{k})=
\sqrt{\xi^2(\mathbf{k})+ \Delta^2(\mathbf{k})  +
\Delta_{d_{xy}}^2}.
\end{equation}
Note that the same result holds for $i s$ order. Making an
integration by parts one may check that the superfluid density
$\rho_s$ remains finite when $\Delta(\mathbf{k})=0$, because the
second gap is also superconducting and $\rho_s =0$ only when both
gaps are zero, $\Delta(\mathbf{k})=\Delta_{d_{xy}}=0$.

For the competing SDW order
\begin{equation}\label{rho_n-SDW}
\Lambda_n^{ij}  = \int \frac{d^2 k}{(2 \pi)^2} v_{Fi}(\mathbf{k})
v_{Fj}(\mathbf{k}) \left[ \frac{2 m^2}{E^3(\mathbf{k})} \tanh
\frac{E(\mathbf{k})}{2T} + \frac{\xi^2(\mathbf{k})
+\Delta^{2}(\mathbf{k})}{T E^2(\mathbf{k})} \frac{1}{\cosh^2
\frac{E(\mathbf{k})}{2T}}\right],
\end{equation}
It is easy to check that  when $\Delta(\mathbf{k}) =0$, while $m$
remains finite, the superfluid density $\rho_s$ becomes zero. In
contrast to the superfluid density, the thermal conductivity is
blind with respect to quantum numbers distinguishing the gaps $m$
and $\Delta_{d_{xy}}$ ($\Delta_s$), so that Eq.~(\ref{thermal-VP})
is valid for all competing SDW, $i s$ and $i d_{xy}$ orders.

Using the nodal approximation \cite{Durst2000PRB} one can estimate
that the depletion of the condensate caused by developing SDW order
at $T=0$
\begin{equation}
\label{Pi-clean-SDW-nodal} \Lambda_n = \frac{v_F}{\pi v_{\Delta}}
|m| \approx N(0) v_F^2 \frac{|m|}{\Delta},
\end{equation}
where $N(0)$ is the density of states (DOS) per  spin in the normal
state and $\Delta$ is the amplitude of the $d$-wave gap.

\section{Influence of the secondary gap opening on the impurity scattering rate}
\label{sec:impurity}

In this section we consider the influence of nonmagnetic impurities
on the residual scattering rate in the presence of a competing
order. We begin with the Hamiltonian written in Nambu formalism
\begin{equation}
H_{imp} = \int d \mathbf{k} \int d \mathbf{k}^\prime
V_{\mathbf{k},\mathbf{k}^\prime} \Psi^\dagger(\tau,\mathbf{k})
\tau_3 \Psi(\tau,\mathbf{k}^\prime),
\end{equation}
describing the interaction $V(\mathbf{r}) = \sum_i V
\delta(\mathbf{r}-\mathbf{r}_i)$ with $\mathbf{r}_i$ the positions
of a random distribution of impurities. Accordingly in the $4
\times 4$ formalism
\begin{equation}
\begin{split}
H_{imp} & = \int_{\mathrm{HBZ}} d \mathbf{k} \int_{\mathrm{HBZ}} d
\mathbf{k}^\prime V_{\mathbf{k},\mathbf{k}^\prime}
\Psi^\dagger(\tau,\mathbf{k}) \hat{I}_2 \otimes \tau_3
\Psi(\tau,\mathbf{k}^\prime)\\& = \int_{\mathrm{HBZ}} d \mathbf{k}
\int_{\mathrm{HBZ}} d \mathbf{k}^\prime
V_{\mathbf{k},\mathbf{k}^\prime} \bar{\Psi}(\tau,\mathbf{k})
(-\gamma_5 \gamma_1) \Psi(\tau,\mathbf{k}^\prime).
\end{split}
\end{equation}
Therefore the set of equations for the impurity self-energy reads
\begin{subequations}
\begin{align}
 G_{0}^{-1}(i\omega,\mathbf{k}) = i\omega \gamma_0 -
\xi(\mathbf{k}) \gamma_1 -
\Delta(\mathbf{k}) \gamma_2 - m \hat{I}_4 - \{\Delta_{d_{xy}} \gamma_3\},\\
 G^{-1}(i {\tilde \omega}, \bk) = G_{0}^{-1}(i {\tilde \omega},\bk)
-
\Sigma (i {\tilde \omega}),\\
 \Sigma (i {\tilde \omega}) =  \Gamma_{imp}  (-\gamma_5
\gamma_1) [c+
g(i {\tilde \omega})\gamma_5 \gamma_1]^{-1}, \label{Sigma_imp}\\
 g(i {\tilde \omega}) = \frac{1}{\pi N(0)} \int \frac{d^2 k}{(2
\pi)^2}G(i {\tilde \omega}, \bk) \label{g},
\end{align}
\end{subequations}
where $c = 1/(\pi N(0)V)$ is the parameter which controls the
strength of impurity scattering and $\Gamma_{imp} = n_{imp}/(\pi
N(0))$ with $n_{imp}$ being the concentration of impurities. Since
here we used an unexpanded dispersion $\xi(\mathbf{k})$ and
$d$-wave gap $\Delta(\mathbf{k})$, the integration in
Eq.~(\ref{g}) is done over the whole Brillouin zone to count
correctly the contribution from all nodes.

Expanding the self-energy $\Sigma$  and $g(i {\tilde \omega})$ in
the $\gamma$-matrices
\begin{equation}
\label{Sigma&g-decomposed}
\begin{split}
& \Sigma (i {\tilde \omega}) = \Sigma^0 (i {\tilde \omega})
\gamma_0 + \Sigma^2(i {\tilde \omega}) \gamma_2 + \Sigma^I(i
{\tilde \omega}) \hat{I}_4 + \{\Sigma^3(i {\tilde \omega})
\gamma_3\} , \\
& g (i {\tilde \omega}) = g^0 (i {\tilde \omega}) \gamma_0 + g^2(i
{\tilde \omega}) \gamma_2  + g^I(i {\tilde \omega}) \hat{I}_4 +
\{g^3(i {\tilde \omega}) \gamma_3\} ,
\end{split}
\end{equation}
one can obtain the system of self-consistent equations for
\begin{equation}\label{Sigma2tilde}
\begin{split}
i {\tilde \omega} & = i \omega - \Sigma^0 (i {\tilde \omega}),\\
{\tilde \Delta}(\bk,i{\tilde \omega} ) & = \Delta(\bk)  +
\Sigma^2 (i {\tilde \omega}),\\
{\tilde m}(i{\tilde \omega} ) & = m  +
\Sigma^I (i {\tilde \omega}),\\
\biggl\{ {\tilde \Delta_{d_{xy}}}(\bk,i{\tilde \omega} ) & =
\Delta_{d_{xy}}(\bk) + \Sigma^3 (i {\tilde \omega}) \biggr\}.
\end{split}
\end{equation}
Assuming particle-hole symmetry for simplicity, the renormalization
of $\xi$ is zero. In particular, when the competing orders are
absent, the system of equations reduces to $T$-matrix equations for
$d$-wave superconductor studied, for example, in
Ref.~\cite{Balatsky1995PRB}. Since in this case the averaging over
the Fermi surface gives $g^2(i {\tilde \omega})=0$, the only
relevant equation left is for $\tilde \omega$ or $\Sigma^0$.

Here our goal is to take into account the influence of a competing
order on $\tilde \omega$. When a competing order develops, it
affects both the  equation mentioned above for $\tilde \omega$
($\Sigma^0$) and the new equation for $\Sigma^I (i {\tilde \omega})$
$\{ \Sigma^3 (i {\tilde \omega})\}$. However, because we do have an
explicit gap equation for $m$ $\{\Delta_{d_{xy}}\}$ (see e.g.
Ref.~\cite{Gorbar1994SC}), in what follows we will assume that the
dependence $\tilde m(\tilde \omega (\omega=0))$ is given
phenomenologically and do not consider an equation for $\Sigma^I (i
{\tilde \omega})$ $\{ \Sigma^3 (i {\tilde \omega})\}$. This
assumption means that we do not distinguish the values $\tilde
m(\tilde \omega (\omega=0))$  and $m$, so that in what follows we
denote the competing gaps as $m$, $\Delta_{d_{xy}}$ and $\Delta_s$.
These gaps already include the effects of impurities and magnetic
field and correspond to their phenomenological values extracted from
experiment. Although in what follows we do not consider the equation
for $\Sigma^I (i {\tilde \omega})$, it is useful to stress the
analogy between the the effect of nonmagnetic impurities on SDW
order, and magnetic impurities in conventional $s$-wave
superconductors \cite{Maki-Parks} that lead to the finite density of
states inside the gap. Physically this means that the scattering off
random impurities prefers to make the system homogeneous by washing
out the nonuniform SDW structure \cite{Dolgov2002PRB}.

Two other important assumptions that we make are the following:\\
(i) We are interested in the value of the zero energy impurity
scattering rate, $\Gamma = -\mbox{Im} \Sigma^0_R(\tilde\omega(\omega
\to 0))$ and do not consider the effects related to the energy
dependence of the self-energy $\Sigma^0(\tilde\omega(\omega))$
studied in Ref.~\cite{Kim2003PRB}. This assumption is justified by
the fact that LSCO compound is intrinsically more dirty system than
YBCO considered in
Ref.~\cite{Kim2003PRB}. \\
(ii) Although we assume that the values of the competing gaps are
field dependent, e.g. $m= m(H)$ we do not include the influence of
the Doppler shift on $\Gamma$ \cite{Kim2003PRB,Kim2004JSC}.

\subsection{$T$-matrix equation for competing SDW order}

We begin with the equation for $\Sigma^0$ for competing SDW order
\begin{equation} \label{SDW-Tmatrix-omega}
\begin{split}
\Sigma^0_{R} ({\tilde \omega})  = \frac{\Gamma_{imp}}{4} & \left[
\frac{1}{c-g^0({\tilde \omega})-g^{I}({\tilde \omega})}+
\frac{1}{-c-g^0({\tilde \omega})+g^{I}({\tilde \omega})} \right.\\
&\left. +\frac{1}{c-g^0({\tilde \omega})+g^{I}({\tilde
\omega})}-\frac{1}{c+g^0({\tilde \omega})+g^{I}({\tilde \omega})}
\right]
\end{split}
\end{equation}
which is obtained from Eq.~(\ref{Sigma_imp}) with $g(i \tilde \omega
\to \tilde \omega)$ given by Eq.~(\ref{g}) using the decompositions
(\ref{Sigma2tilde}).

Since we are interested in the value $\Sigma^0(\tilde\omega)$ with
$\tilde \omega = \omega - \Sigma^0(\tilde \omega)$ at $\omega = 0$,
we need only the functions $g^0(\tilde \omega)$ and $g^I(\tilde
\omega)$ calculated for $\tilde \omega = i \Gamma$. Using the nodal
approximation we obtain
\begin{subequations}\label{g0-gI}
\begin{align}
g^0(i \Gamma) = \frac{1}{\pi N(0)} \frac{1}{\pi v_F v_{\Delta}}
\ln \frac{m^2 + \Gamma^2}{p_0^2} i \Gamma, \label{g0} \\
g^I(i \Gamma) = \frac{1}{\pi N(0)} \frac{1}{\pi v_F v_{\Delta}} \ln
\frac{m^2 + \Gamma^2}{p_0^2} m, \label{gI}
\end{align}
\end{subequations}
where the ultraviolet cutoff $p_0$ is introduced. Note that for $p_0
\sim 4 \Delta$ and $(v_{F} v_{\Delta})^{-1} \sim \pi N(0)/\Delta$,
where $\Delta$ is the amplitude of the $d$-wave gap $\Delta(\bk) =
\Delta/2 (\cos k_x a - \cos k_y a)$, the function $g(\omega)$
calculated using the nodal approximation agrees with the expressions
given in Refs.~\cite{Balatsky1995PRB,Kim2003PRB,Kim2004JSC}. In
Eq.~(\ref{SDW-Tmatrix-omega}) the Born limit corresponds to $V \to
0$ ($c \to \infty$), while the unitary limit corresponds to $V \to
\infty$, i.e. $c \to 0$.

\subsubsection{Born limit}
\label{sec:Born-SDW}

In the Born limit Eq.~(\ref{SDW-Tmatrix-omega}) reduces to
\begin{equation}
\label{SDW-Born} \Sigma^0_R ({\tilde \omega})  = \Gamma_{imp} c^{-2}
g^0({\tilde \omega}).
\end{equation}
Due to the fact that only $g^0$ enters Eq.~(\ref{SDW-Born}), the
Born limit appears to be the same both for SDW and superconducting
orders, because $g^0$ depends either on $m^2$ or on
$\Delta_{{d}_{xy}}^2$.

Substituting Eq.~(\ref{g0}) in Eq.~(\ref{SDW-Born}) and solving it
with respect to $\Gamma$ we obtain \cite{Franz2001PRB}
\begin{equation}\label{Gamma-eq-Born}
\Gamma^2 = p_0^2 \exp \left[-\frac{c^2 \pi^2 v_F v_\Delta
N(0)}{\Gamma_{imp}}\right] -m^2.
\end{equation}
For a given impurity concentration this solution is nonzero only if
$m< m_{cr}$, where $m_{cr} = p_0 \exp [-c^2 \pi^2 v_F v_\Delta
N(0)/(2\Gamma_{imp})]$. Because the Born limit is considered, the
value of $m_{cr}$ is exponentially small. Nevertheless in the
unitary limit there is a possibility to have both finite $\Gamma$
and rather large values of $m$.

\subsubsection{Unitary limit}
\label{sec:unitary-SDW}

\begin{figure}[h]
\centering{
\includegraphics[width=8cm]{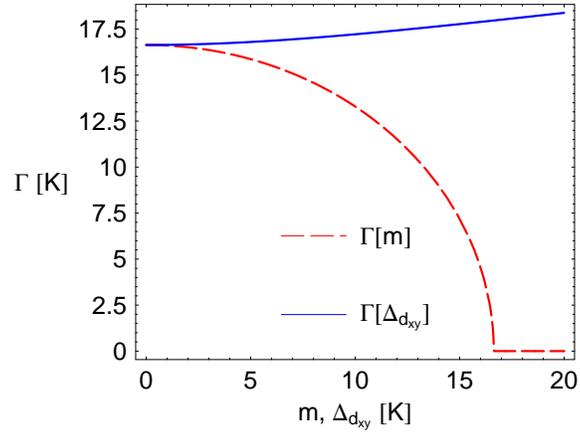}}
\caption{(Color online) The dependence of the impurity scattering
rate $\Gamma$ on the value of the gaps $m$ and $\Delta_{d_{xy}}$ of
the developing SDW and $i d_{xy}$ orders, respectively. For
$m>m_{ct}$ the value $\Gamma(m)=0$.} \label{fig:2}
\end{figure}

LSCO compound is intrinsically a dirtier system than other cuprates,
so that the unitary limit is more relevant. Moreover, thin LSCO
films studied in Ref.~\cite{Martinoli.report} seems to have
particularly large values of $\Gamma \sim 6 \div 50 \mbox{K}$ which
also indicates the relevance of the unitary limit. When $c \to 0$
Eq.~(\ref{SDW-Tmatrix-omega}) reduces to
\begin{equation}\label{SDW-unitary}
\Sigma^0_R ({\tilde \omega}) = \Gamma_{imp} \frac{g^0({\tilde
\omega})}{[g^I({\tilde \omega})]^2 - [g^0({\tilde \omega})]^2}.
\end{equation}
Now because $g^I$ still enters Eq.~(\ref{SDW-unitary}), the
resulting transcendental equation for $\Gamma$ is
\begin{equation}
\label{SDW-unitary-eq} \Gamma^2 =  \Gamma_{imp} \pi^2 N(0) v_F
v_\Delta \left[ \ln \frac{p_0^2}{\Gamma^2 + m^2}\right]^{-1}- m^2,
\end{equation}
which differs from that of Ref.~\cite{Franz2001PRB} by the last term
$m^2$. Due to its presence the dependence $\Gamma(m)$ is rather
strong (see Fig.~\ref{fig:2}) and as in the Born limit there is a
critical value $m_{cr}^{\mathrm{unit}}$ such that $\Gamma(m)=0$ for
$m>m_{cr}^{\mathrm{unit}}$. As we will argue below the observed
deviations of the dependence $\rho_s(H)$ (or $\Lambda_s(H)$) from
$\sim \sqrt{H}$ in high fields can be caused by the fact that the
field dependence of $m(H)$ affects the behavior of $\rho_s(H)$ via
the dependence $\Gamma(m(H))$. Finally we note that the case of
competing $i s$ order is also described by
Eq.~(\ref{SDW-unitary-eq}).

\subsection{$T$-matrix equation for competing $i d_{xy}$ order}

Similarly to  Eq.~(\ref{SDW-Tmatrix-omega}) for the competing $i
d_{xy}$ order we obtain
\begin{equation}
\label{SC-Tmatrix} \Sigma^0_R ({\tilde \omega})  = \Gamma_{imp}
\frac{g^0({\tilde \omega})}{c^2 -[g^0({\tilde \omega})]^2 +
[g^3({\tilde \omega})]^2},
\end{equation}
where $g^0(i \Gamma)$ is given by  (\ref{g0}), while for $d_{xy}$
order $g^3 =0$, analogously to the function $g^2$ for $d_{x^2 -y^2}$
case.

\subsubsection{Born limit}
\label{sec:Born-SC}

As was already mentioned in Sec.~\ref{sec:Born-SDW}, in the Born
limit there is no difference between the consideration of competing
SDW and $i d_{xy}$ order, so that one may simply replace the gap $m$
by $\Delta_{d_{xy}}$ in the corresponding equations. Moreover, this
consideration is also valid for a competing $is$ order.  Competing
$is$ order with $d$-wave superconductivity  was considered in
Ref.~\cite{Shurrer1998PC}, where besides the $T$-matrix equations
for the impurity scattering rate, the optical conductivity  order
was studied.

\subsubsection{Unitary limit}
\label{sec:unitary-SC}

Since $g^3=0$, instead of Eq.~(\ref{SDW-unitary-eq}) we arrive at
the equation \cite{Franz2001PRB}
\begin{equation}
\label{SC-unitary-eq} \Gamma^2 =  \Gamma_{imp} \pi^2 N(0) v_F
v_\Delta \left[ \ln \frac{p_0^2}{\Gamma^2 +
\Delta_{d_{xy}}^2}\right]^{-1}.
\end{equation}
Its solution $\Gamma(\Delta_{d_{xy}})$ is shown in Fig.~\ref{fig:2}.
It demonstrates that a competing $i d_{xy}$ does not significantly
perturb the value of $\Gamma$ with respect to the
$\Delta_{d_{xy}}=0$ case, when it reduces to the form
\cite{Vekhter2001PRB}
\begin{equation}
\Gamma^2 = \frac{\pi}{2} n_{imp} v_F v_\Delta \left[ \ln
\frac{p_0}{\Gamma}\right]^{-1}.
\end{equation}

\section{Superfluid density for $H=0$ }
\label{sec:superfluid_gap}

In Sec.~\ref{sec:rho_s-formal} we discussed in a more formal way how
the different competing orders affect the superfluid density. Here
instead we concentrate on the simple analytical expressions that
demonstrate the dependence of $\Lambda_n$ on the impurity scattering
rate $\Gamma$ and the competing gaps $m$ and $\Delta_{d_{xy}}$. In
the nodal approximation  the representation (\ref{rho.n-RA})
acquires a form convenient for analytical calculations
\begin{equation}
\label{Lambda2J} \Lambda_n = - \frac{v_F }{2 \pi v_\Delta} J,
\end{equation}
where
\begin{equation}
\label{J2I} J= - \int_{0}^\infty d \omega \tanh \frac{\omega}{2T}
\tilde{I} (\omega)
\end{equation}
with
\begin{equation}
\tilde{I} (\omega)=  \frac{1}{2 \pi i} \int_0^{p_0} p d p
[I_{A}(\omega, p)- I_{R}(\omega, p)]
\end{equation}
and
\begin{equation}
\label{IRA} I_{R,A}(\omega, p) \equiv  \mbox{tr}[ G_{R,A}(\omega, p)
\gamma_0 \gamma_5 G_{R,A}(\omega,p)\gamma_0 \gamma_5].
\end{equation}
Here $p = \sqrt{v_F^2 k_1^2 + v_{\Delta}^2 k_2^2}$ is the dispersion
law of the quasiparticles in the nodal approximation. $J$ in
Eq.~(\ref{Lambda2J}) is twice bigger than in
Ref.~\cite{Sharapov2002PRB} because a $4 \times4$ formalism is used.
Now we consider the cases of competing SDW, $i d_{xy}$ and $i s$
orders.

\subsection{Competing SDW order}

Substituting the Green's function (\ref{GF_I}) with the self-energy
$\mbox{Im} \Sigma_{R,A}(\omega =0) = \mp \Gamma$ in Eq.~(\ref{IRA})
one obtains
\begin{equation}
\label{IRA-SDW} I_{R,A}(\omega, p) = \frac{4[(\omega \pm i \Gamma)^2
+ p^2 -m^2]}{[(\omega \pm i \Gamma)^2 - p^2- m^2]^2}.
\end{equation}
Then integration over the energy $p$ we arrive at
\begin{equation}\label{I_m}
\begin{split}
\tilde{I} (\omega)=  - \frac{2}{\pi} &\left[\arctan
\frac{p_0^2+\Gamma^2+m^2 -\omega^2}{2\Gamma\omega}
- \arctan \frac{\Gamma^2+m^2 -\omega^2}{2\Gamma\omega} \right.\\
&\left. - \frac{p_0^2 \Gamma}{\sqrt{p_0^2 +
m^2}}\frac{1}{(\omega-\sqrt{p_0^2 + m^2})^2 + \Gamma^2} +
\frac{p_0^2 \Gamma}{\sqrt{p_0^2 + m^2}}\frac{1}{(\omega+\sqrt{p_0^2
+ m^2})^2 + \Gamma^2} \right].
\end{split}
\end{equation}
Finally integrating over $\omega$ for $\Gamma \gg T$ (also for $p_0
\gg m,\Gamma$) we obtain
\begin{equation}\label{rho_n-SDW-final}
\Lambda_n =  \frac{v_F}{\pi^2 v_{\Delta}} \left[ \Gamma \ln
\frac{p_0^2}{m^2 + \Gamma^2} + m \arctan \frac{2 m \Gamma}{\Gamma^2
- m^2} + \pi |m| \theta(m^2 - \Gamma^2)+ \frac{\pi^2}{3} \frac{T^2
\Gamma}{\Gamma^2+m^2}\right],
\end{equation}
where $\theta$ is the step function. The first term of
Eq.~(\ref{rho_n-SDW-final}) can be interpreted as the DOS
contribution to the depletion of the condensate even at $T=0$,
because the DOS (per spin) in a dirty $d$-wave superconductor with a
finite gap $m$ reads \cite{Gusynin2004EPJB}
\begin{equation}
N_m(0) = \frac{2}{\pi^2 v_F v_{\Delta}} \Gamma \ln
\frac{p_0}{\sqrt{\Gamma^2 + m^2}}.
\end{equation}
The second and third terms of Eq.~(\ref{rho_n-SDW-final}) describe
the depletion of the condensate because of the development of SDW
order and for $\Gamma \to 0$ they reduce to
Eq.~(\ref{Pi-clean-SDW-nodal}) discussed above. Finally the last
term of Eq.~(\ref{rho_n-SDW-final}) $\sim T^2$ shows that the
characteristic for $d$-wave superconductors i.e. a linear $T$
dependence for $\Lambda_n$ changes in the presence of impurities and
becomes $\sim T^2$. This behavior is indeed observed in thin films
\cite{Martinoli.report} over a wide range of the temperatures. Based
on Eq.~(\ref{rho.dirty.final2}) below, in some films it was
estimated that $\Gamma$ can be as big as $ \approx 46.6 \mbox{K}$
\cite{Martinoli.report}. When $m\to 0$, Eq.~(\ref{rho_n-SDW-final})
reduces to the known
\cite{Hirschfeld1993PRB,Xiang1998IJMPB,Sharapov2002PRB} expression
\begin{equation}\label{rho.dirty.final2}
\Lambda_{n}(T) = \frac{v_{F}}{\pi v_{\Delta}} \left[\frac{2
\Gamma}{\pi} \ln \frac{p_0}{\Gamma} + \frac{\pi}{3}
\frac{T^2}{\Gamma} - O\left(\frac{T^4}{\Gamma^3}\right) \right].
\end{equation}
Again the first term of Eq.~(\ref{rho.dirty.final2}) is proportional
to the DOS of a dirty $d$-wave superconductor \cite{Durst2000PRB}.

It is also instructive to consider the limit $\Gamma \to0$ when the
$T$-dependence of $\Lambda_n$ is expected to change from quadratic
to linear. To extract this limit one should extract the singular
part of $\tilde{I} (\omega)$ in the limit $\Gamma \to 0$ , viz.
\begin{equation}
\label{ISDW-clean} \tilde{I}(\omega) = -\mbox{sgn} \omega
[\mbox{sgn}(p_0^2 - \omega^2) - \mbox{sgn}(m^2 - \omega^2)] + 2 p_0
[\delta(\omega-p_0)-\delta(\omega+p_0)],
\end{equation}
where we again took $p_0 \gg m,\Gamma$. Then
\begin{equation}
\label{J-SDW-clean} J(m) = - 4 T \ln\left( 2\cosh
\frac{m}{2T}\right).
\end{equation}
Obviously for $m= 0$ we recover the well-known expression
\cite{Durst2000PRB}
\begin{equation}
\label{rho_n-d-wave} \Lambda_n= \frac{2 \ln2}{\pi}
\frac{v_F}{v_\Delta} T.
\end{equation}
For $|m| \gg T$ we obtain
\begin{equation}
\label{rho_n-SDW-exp} \Lambda_n = \frac{v_F}{\pi v_{\Delta}} [|m| +
2 T \exp(-|m|/T)],
\end{equation}
where the first term coincides with Eq.~(\ref{Pi-clean-SDW-nodal}).
For $|m| \ll T$, we arrive at the expression
\begin{equation}
\label{rho_n-SDW-asymp} \Lambda_n = \frac{2 \ln2}{\pi}
\frac{v_F}{v_\Delta} T \left[ 1+ \frac{1}{8 \ln2}
\frac{m^2}{T^2}\right].
\end{equation}
Here we  refer for comparison to simple expressions for $\Lambda_n$
when $d$-wave superconducting states coexists with the orbital
antiferromagnetic ($d$-density-wave) state \cite{Kee2002PRB}.

\subsection{Competing $i d_{xy}$ and $is$ orders}

The difference between competing SDW and superconducting orders can
be traced back to the opposite sign before the $m^2$ term in the
numerator of Eq.~(\ref{IRA-SDW}) and the corresponding sign before
$\Delta_{d_{xy}}^2$ in the formula
\begin{equation}
\label{IRA-SC} I_{R,A}(\omega, p)  = \frac{4[(\omega \pm i \Gamma)^2
+ p^2 +\Delta_{d_{xy}}^2]}{[(\omega \pm i \Gamma)^2 - p^2-
\Delta_{d_{xy}}^2]^2}
\end{equation}
which is related to the different $\gamma$-matrix before
$\Delta_{d_{xy}}$ in Eq.~(\ref{GF_id}). Note that for competing $is$
order, Eq.~(\ref{IRA-SC}) turns out to be the same, so that these
order differ by the influence of the  superconducting gaps
$\Delta_{d_{xy}}$ and $\Delta_s$ on the impurity scattering rate
$\Gamma$ (see Sec.~\ref{sec:unitary-SDW}) in the unitary limit.
Integrating over $p$ one obtains
\begin{equation}\label{I_SC}
\begin{split}
\tilde{I} (\omega)=  - \frac{2}{\pi} &\left[\arctan
\frac{p_0^2+\Gamma^2+\Delta_{d_{xy}}^2 -\omega^2}{2\Gamma\omega} -
\arctan \frac{\Gamma^2+\Delta_{d_{xy}}^2 -\omega^2}{2\Gamma\omega}\right.\\
& \left.+ \frac{\Gamma\Delta_{d_{xy}}}{(\omega-\Delta_{d_{xy}})^2 +
\Gamma^2} -\frac{\Gamma\Delta_{d_{xy}}}{(\omega+\Delta_{d_{xy}})^2 +
\Gamma^2}
\right.\\
&\left. - \frac{\Gamma \sqrt{p_0^2 +
\Delta_{d_{xy}}^2}}{(\omega-\sqrt{p_0^2 + \Delta_{d_{xy}}^2})^2 +
\Gamma^2} + \frac{\Gamma \sqrt{p_0^2 +
\Delta_{d_{xy}}^2}}{(\omega+\sqrt{p_0^2 + \Delta_{d_{xy}}^2})^2 +
\Gamma^2} \right].
\end{split}
\end{equation}
Finally integrating over $\omega$ for $\Gamma \gg T$ (also for $p_0
\gg \Delta_{d_{xy}},\Gamma$) we obtain
\begin{equation}\label{rho_n-SC-final}
\Lambda_n =  \frac{v_F}{\pi^2 v_{\Delta}} \left[ \Gamma \ln
\frac{p_0^2}{\Delta_{d_{xy}}^2 + \Gamma^2} + \frac{\pi^2}{3}
\left(\frac{\Gamma}{\Gamma^2+\Delta_{d_{xy}}^2}+ \frac{2
\Delta_{d_{xy}}^2 \Gamma}{(\Gamma^2+\Delta_{d_{xy}}^2)^2}\right)
T^2\right].
\end{equation}

Similarly to Eq.~(\ref{ISDW-clean}) in the limit $\Gamma \to0$ from
Eq.~(\ref{I_SC}) we obtain
\begin{equation}
\label{ISC-clean}
\begin{split}
\tilde{I} (\omega)= & - \mbox{sgn} \omega [\mbox{sgn}(p_0^2
-\omega^2) - \mbox{sgn}(\Delta_{d_{xy}}^2 -\omega^2)]\\
&+ 2 p_0 [\delta(\omega-p_0)- \delta(\omega+ p_0)] - 2
\Delta_{d_{xy}} [\delta(\omega- \Delta_{d_{xy}})- \delta(\omega+
\Delta_{d_{xy}})],
\end{split}
\end{equation}
where we again took $p_0 \gg \Delta_{d_{xy}},\Gamma$. Then
\begin{equation}
J(\Delta_{d_{xy}}) = 2 \Delta_{d_{xy}} \tanh
\frac{\Delta_{d_{xy}}}{2T} - 4 T \ln \left(2 \cosh
\frac{\Delta_{d_{xy}}}{2T}\right),
\end{equation}
which differs from Eq.~(\ref{J-SDW-clean}) by the term with $\tanh$.
For $\Delta_{d_{xy}}= 0$ we again recover  Eq.~(\ref{rho_n-d-wave}).
For $\Delta_{d_{xy}} \gg T$ the dependence $\Lambda_n(T)$ becomes
thermally activated due to the secondary gap
\begin{equation}
\label{rho_n-SC-exp} \Lambda_n = \frac{2}{\pi} \frac{v_F}{v_\Delta}
\Delta_{d_{xy}} \exp(-\Delta_{d_{xy}}/T),
\end{equation}
while for $\Delta_{d_{xy}} \ll T$ the leading term of $\Lambda_n(T)$
coincides with Eq.~(\ref{rho_n-d-wave})
\begin{equation}
\label{rho_n-SC-asymp} \Lambda_n = \frac{2 \ln2}{\pi}
\frac{v_F}{v_\Delta} T \left[ 1- \frac{1}{8 \ln2}
\frac{\Delta_{d_{xy}}^2}{T^2}\right].
\end{equation}
Comparing the last equation with Eq.~(\ref{rho_n-SDW-asymp}) one
may notice that they differ by the sign before the $T^2$ term.

\section{Superfluid density in the vortex state}
\label{sec:superfluid-vortex}

The presence of circulating supercurrent around vortices in the
mixed state can be taken into account in the semiclassical approach
by introducing the Doppler shift in quasiparticle energies, $\omega
\to \omega- \mathbf{v}_s(\mathbf{r}) \mathbf{k}$
\cite{Volovik1993JETPL}. Here $\mathbf{v}_s(\mathbf{r})$ is the
superfluid velocity at a position $\mathbf{r}$ which depends on the
form of the vortex distribution and $\mathbf{k}$ is the
quasiparticle momentum which can be approximated by its value at the
node \cite{Vekhter2001PRB}. This distribution is described by the
function
\begin{equation}
\mathcal{P}(\epsilon) = \frac{1}{A} \int d^2 r \delta(\epsilon -
\mathbf{v}_s (\mathbf{r}) \mathbf{k}),
\end{equation}
where the integration is over the unit vortex cell with the area
$A = \pi R^2$. Several choices for $\mathcal{P}(\epsilon)$ were
discussed in Ref.~\cite{Vekhter2001PRB}. Among them  are the
distribution for the vortex liquid \cite{Vekhter2001PRB}
\begin{equation}
\label{P-liquid} \mathcal{P}(\epsilon) =  \frac{E_H^2}{2(\epsilon^2
+ E_H^2)^{3/2}}
\end{equation}
which is the most convenient for analytic calculations, and
\cite{Yu1995PRL,Vekhter2001PRB}
\begin{equation}
\label{P-disorder}  \mathcal{P}(\epsilon) = \frac{1}{\sqrt{\pi} E_H}
\exp\left(-\frac{\epsilon^2}{E_H^2} \right)
\end{equation}
for the completely disordered vortex state. The characteristic
energy scale $E_H$ in Eqs.~(\ref{P-liquid}) and (\ref{P-disorder})
is associated with the Doppler shift energy in the vortex state
\begin{equation}
E_H (H) = a \frac{\hbar v_F}{2R} = a \frac{\hbar v_F}{2} \sqrt{
\frac{\pi H}{\Phi_0}},
\end{equation}
where $a$ is a geometrical factor of order unity and $H$ is the
magnetic field applied perpendicularly to the $ab$ plane. In the
second equality we used the convention of Ref.~\cite{Vekhter2001PRB}
(see also Ref.~\cite{Wang2001PRB} and Ref.~[33] therein) that for
$a=1$ there is one flux quantum $\Phi_0 = h c/2 e$ per unit cell of
the vortex lattice approximated by a circle of radius $R=
(\Phi_0/\pi H)^{1/2}$.
The final results  depend somewhat on the choice of the distribution
function and on the value of $a$, however, the qualitative results
are not sensitive to this choice. In what follows we take the value
$v_F = 2.5 \times 10^7 \mbox{cm/s}$ \cite{Sutherland2003PRB} which
corresponds to $E_{H}[\mbox{K}] =38 \mbox{K} \cdot \mbox{T}^{-1/2}
\sqrt{H [\mbox{T}]}$ as used in Ref.~\cite{Gusynin2004EPJB}.

Now the Doppler shift effect can be incorporated in the Green's
function formalism that already includes the scattering on
impurities by averaging over the distribution
$\mathcal{P}(\epsilon)$ \cite{Kubert1998SSC,Vekhter2001PRB}
\begin{equation} \label{rho_n-H}
\begin{split}
& \Lambda_{n}^{ij}(H) =
  \int_\infty^\infty d
\epsilon \mathcal{P}(\epsilon) \int_{\mathrm{HBZ}} \frac{d^2 k}{(2
\pi)^2} \int_{-\infty}^{\infty} d \omega  \tanh\frac{\omega}{2T}
v_{F i} v_{F j} \frac{1}{4 \pi i}
\\
& \times \mbox{tr}[ G_{A}(\omega-\epsilon, \mathbf{k}) \gamma_0
\gamma_5 G_{A}(\omega-\epsilon,\mathbf{k})\gamma_0 \gamma_5 -
G_{R}(\omega-\epsilon, \mathbf{k}) \gamma_0 \gamma_5
G_{R}(\omega-\epsilon, \mathbf{k})\gamma_0 \gamma_5].
\end{split}
\end{equation}
We note that in $2\times2$ Nambu formalism the replacement $\omega
\to \omega - \epsilon$ for taking into account the Doppler shift
in the argument of the Green's function is exact. In $4 \times 4$
formalism the corresponding Green's function describes two
opposite nodes with the reversed sign of the Doppler shift.
Nevertheless one can check that in the latter case the simple
prescription $\omega \to \omega - \epsilon$ is approximately valid
if we neglect the terms $\sim \epsilon m$. In the nodal
approximation similarly to Eq.~(\ref{Lambda2J}) one may write
\begin{equation}\label{rho_n-final-H}
\Lambda_{n}(H) = - \frac{v_F}{2 \pi v_\Delta}\int_\infty^\infty d
\epsilon \mathcal{P}(\epsilon) J(m(H);\epsilon),
\end{equation}
where the function $J(\epsilon; m(H))$ is given below. However, the
field dependence of $\Lambda_{n}(H)$ is not yet completely
specified, because one should also provide a field dependence for
the gap $m(H)$ $\{ \Delta_{d_{xy}} (H) \}$. We will come to this
question later in Sec.~\ref{sec:ansatz}.

\subsection{Results for the field dependence of the superfluid density for $m, \Delta_{d_{xy}},\Gamma = \mbox{const}$}

For $T=0$ and $p_0 \gg m,\Delta_{d_{xy}}, \Gamma$ in SDW case one
obtains
\begin{equation}
\label{J-SDW-H}
\begin{split}
J(m,T=0;\epsilon)  = -\frac{2}{\pi} & \left[\epsilon \arctan
\frac{\epsilon^2 - m^2 - \Gamma^2}{2 \epsilon \Gamma} +
\frac{\pi}{2} |\epsilon|
\right. \\
& \left. + \Gamma \ln \frac{p_0^2}{\sqrt{[\Gamma^2 +(m -
\epsilon)^2][ \Gamma^2 + (m+\epsilon)^2]}} \right. \\ & \left. + m
\arctan \frac{2 m \Gamma}{\Gamma^2 +\epsilon^2 - m^2} + \pi |m|
\theta(m^2 - \Gamma^2 -\epsilon^2)\right]
\end{split}
\end{equation}
and in the $i d_{xy}$ case
\begin{equation}
\label{J-SC-H}
\begin{split}
J(\Delta_{d_{xy}},T=0;\epsilon)  = -\frac{2}{\pi} & \left[\epsilon
\arctan \frac{\epsilon^2 - \Delta_{d_{xy}}^2 - \Gamma^2}{2 \epsilon
\Gamma} + \frac{\pi}{2} |\epsilon|
\right. \\
& \left. +  \Gamma \ln \frac{p_0^2}{\sqrt{[\Gamma^2
+(\Delta_{d_{xy}} - \epsilon)^2][ \Gamma^2 +
(\Delta_{d_{xy}}+\epsilon)^2]}}\right].
\end{split}
\end{equation}

Before doing the numerical calculation it is useful to consider
analytically the limit $\Gamma \to0$. One may notice that in this
limit the only difference between the SDW and $i d_{xy}$ cases is
the last two terms $\sim m$ in Eq.~(\ref{J-SDW-H}). Thus we
consider firstly the $i d_{xy}$ case and then discuss the role of
these terms. We obtain from Eq.~(\ref{J-SC-H}) that
\begin{equation}
\label{J0-SC-H} J(\Delta_{d_{xy}},T=0;\epsilon)  = -2 |\epsilon|
\theta(\epsilon^2 - \Delta_{d_{xy}}^2).
\end{equation}
Substituting Eq.~(\ref{J-SC-H}) in Eq.~(\ref{rho_n-final-H}) and
using the vortex liquid distribution (\ref{P-liquid}) we arrive at
\begin{equation}
\Lambda_{n}(H) =  \frac{v_F}{\pi v_\Delta} \frac{E_H^2}{\sqrt{E_H^2+
\Delta_{d_{xy}}^2}}.
\end{equation}
This means that in spite of the presence of a gap for nodal
quasiparticles, the contribution to $\Lambda_n$ still has the
behavior $\sim \sqrt{H}$ if $\Delta_{d_{xy}}(H) \lesssim \sqrt{H}$.
The origin of the gapless behavior is the Doppler shift of the
quasiparticle energy $E - \mathbf{k} \mathbf{v}_s (\mathbf{r})$,
which is position dependent. There are regions where the shift is
larger than the minimal gap $\Delta_{d_{xy}}$ in the spectrum, thus
leading to the finite DOS and $\Lambda_n$. This point about the DOS
was emphasized in Refs.~\cite{Hirschfeld1998AMPS,Mao1999PRB}, where
it is stressed that regardless of the power with which
$\Delta_{d_{xy}} \sim H^\alpha$ opens up in the field, as long as
$\alpha \geq 1/2$, the leading term in DOS will always be $\sqrt{H}$
at small fields. Thus in the clean system $\Lambda_{n}(H)$ would
remain $\sim \sqrt{H}$ even in the presence of  a nonzero gap
$\Delta_{d_{xy}} \varpropto H^\alpha$ with $\alpha \geq 1/2$. As we
already mentioned in the SDW case the last two terms of
Eq.~(\ref{J-SDW-H}) contribute in $\Lambda_n$. However, if the SDW
gap $m(H) \varpropto \sqrt{H}$, the $\sqrt{H}$ behavior of
$\Lambda_{n}(H)$ would persist, because the contribution of the last
two terms of Eq.~(\ref{J-SDW-H}) is $v_F/(\pi v_{\Delta})
m^2(H)/\sqrt{E_H^2 + m^2(H)}$.

\begin{figure}[h]
\centering{
\includegraphics[width=8cm]{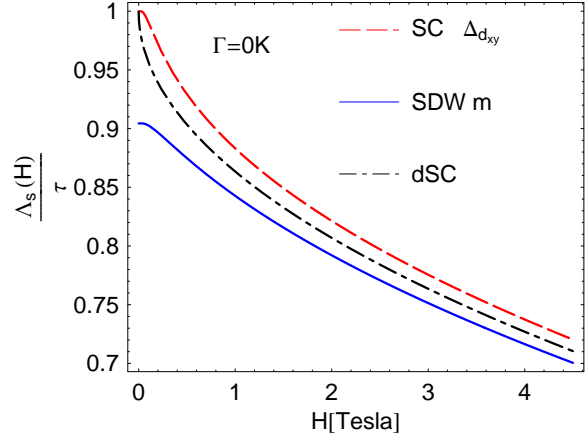}}
\caption{(Color online) The dependence $\Lambda_s(H))/\tau$ at $T=0$
for $d$-wave superconductor with additional SDW, $id_{xy}$ gaps
$\Delta_{d_{xy}}= m = 15 \mbox{K}$ and also without any competing
order in the clean, $\Gamma=0 \mbox{K}$ limit. } \label{fig:3}
\end{figure}

\begin{figure}[h]
\centering{
\includegraphics[width=8cm]{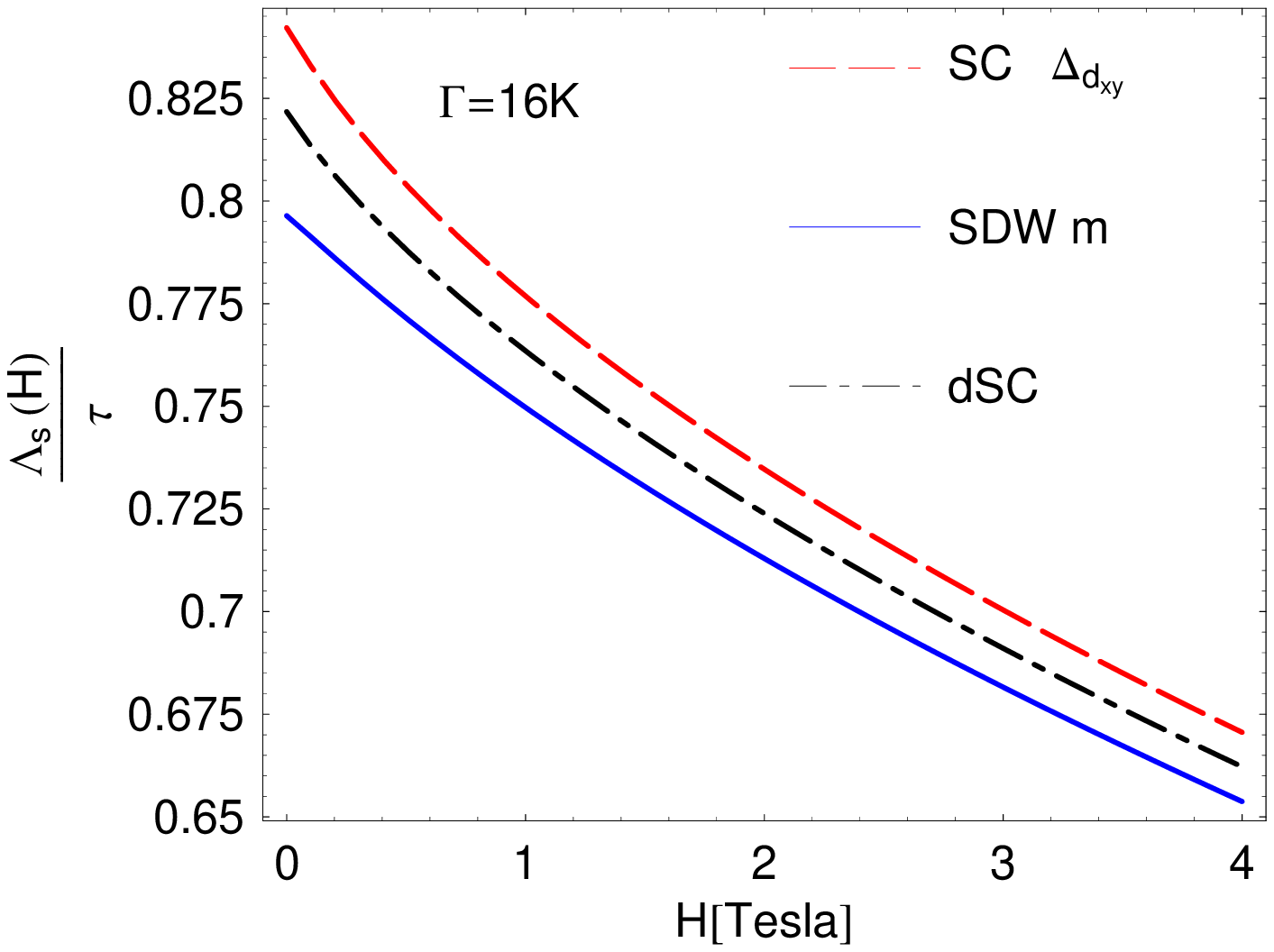}}
\caption{(Color online) The dependence $\Lambda_s(H))/\tau$ at $T=0$
for $d$-wave superconductor with additional SDW, $id_{xy}$ gaps
$\Delta_{d_{xy}}= m = 15 \mbox{K}$ and also without any competing
order. The constant $\Gamma=16 \mbox{K}$ is taken. } \label{fig:4}
\end{figure}

In Figs.~\ref{fig:3} and \ref{fig:4} we plot the dependence
$\Lambda_s(H)$ for constant field independent gaps $m$,
$\Delta_{d_{xy}}$ for the clean case ($\Gamma=0 $) and for
$\Gamma=16 \mbox{K}$, respectively. This dependence is obtained
numerically from Eq.~(\ref{rho_n-final-H}) using the distribution
Eq.~(\ref{P-disorder}). We use $v_{F}/v_{\Delta}=30$ and a rough
estimate of the diamagnetic term $\tau = 1500 \mbox{K}$ from the
Uemura plot. In the clean limit $\Gamma=0$ at zero temperature, the
opening of the gap leaves the condensate unaltered if it is of
superconducting character, but depletes it for the spin density wave
case. The application of an external magnetic field $H$ oriented
perpendicular to the cooper oxide planes creates quasiparticles and
decreases the condensate density as compared with its zero field
value. Comparing with the pure $d_{x^2 -y^2}$ case, the reduction in
condensate for a given value of $H$ is less for an additional $i
d_{xy}$ ($is$) gap and even less for the SDW case. However, because
in this last case the $H=0$ value is already depleted as compared
with the pure $d_{x^2-y^2}$ case, the condensate density remains
lower for all values of $H$ considered. Looking at Fig.~\ref{fig:3}
one may develop the impression that the cases of competing $i
d_{xy}$ and SDW orders are easily distinguishable experimentally.
However, the situation is more complicated because in practice one
considers the dependence $\Lambda_s(H) -\Lambda_s(H=0)$ which can
hardly distinguish different competing orders. Moreover, in all
cases the introduction of impurity (residual) scattering reduces the
condensate density and makes the effect of $H$ on it more similar
(Fig.~\ref{fig:4}).

\subsection{Ansatz for gap $m(H)$}
\label{sec:ansatz}

In general the competing gap and its doping dependence have to be
obtained by solving a self-consistent system of equations for the
main $d$-wave and subdominant gaps (see e.g.
Ref.~\cite{Gorbar1994SC,Micnas2002JPCM}). However, since the origin
of the secondary gap is unknown, here we follow the phenomenological
approach of Ref.~\cite{Gusynin2004EPJB}, where it was assumed that
\begin{equation}
\label{gap-gusynin} m(H,x) = (1-x/0.16)^{1/2} \theta(0.16-x) (m_0 +
b E_H),
\end{equation}
where $x$ is the doping, $m_0$ and $b$ are free parameters.

It was demonstrated in Ref.~\cite{Gusynin2004EPJB} that the
experimental data in Refs.~\cite{Sun2003PRL,Hawthorn2003PRL} can
be qualitatively understood using the gap (\ref{gap-gusynin})
which is generated below a critical doping  $x_{cr}=0.16$  and
increases with magnetic field as $\sqrt{H}$
\cite{Laughlin1998PRL}. As mentioned in the introduction, the
present work is partly motivated by the experiments
\cite{Khaykovich2005PRB,Li2005} made on an $x=0.144$ LSCO sample,
where the SDW gap develops above a critical field $H_0 \approx 3
\mbox{T}$. Since here we do not consider the doping dependence of
the gap but instead are more interested in the effects related to
the critical field, we assume that
\begin{equation}
\label{gap-my} m(H) = b E_H(H-H_0).
\end{equation}
Moreover, while in Ref.\cite{Gusynin2004EPJB} a rather large value
of $b=2.2$ was used, in the present paper we consider the case of
small values of $b=0.17$, so that $m(H) \ll E_H$. The dependence
$m(H)$ is shown in Fig.~\ref{fig:5}, where we also plot the
dependence $\Gamma(H)$ obtained by solving
Eq.~(\ref{SDW-unitary-eq}) for $\Gamma(m)$ with $p_0 = 250 \mbox{K}$
and $\pi^2 \Gamma_{imp} N(0) v_F v_{\Delta} = 1500 \mbox{K}^2$. As
we already mentioned, there is also a direct influence of the
Doppler shift and Andreev scattering on $\Gamma$
\cite{Kim2003PRB,Kim2004JSC}. As shown in Ref.~\cite{Kim2004JSC}, in
the unitary limit the change of $\Gamma$ due to the Doppler shift is
not significant. Nevertheless these effects and particularly the
energy dependence of $\Gamma(H,\omega)$ \cite{Kim2003PRB} will
become important when the value of $m \lesssim
m_{cr}^{\mathrm{unit}}$ and $\Gamma(\omega =0)$ approaches zero.

Finally we note  the fact that the explanation of experimental data
requires a doping dependent gap, which  supports its SDW character.
The generation of a $i d_{xy}$ gap by a magnetic field can
presumably occur at any doping.

\subsection{Results for the field dependence of the superfluid density for
field dependent gaps and $\Gamma$}

\begin{figure}[h]
\centering{
\includegraphics[width=8cm]{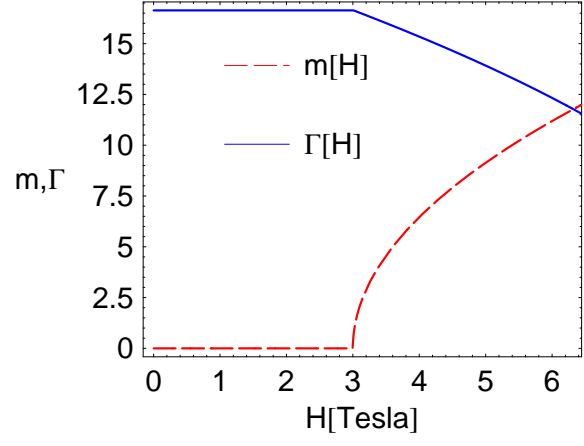}}
\caption{(Color online) The model dependence of the gap $m$ on the
applied field $H$ and the resulting dependence of the impurity
scattering rate $\Gamma$ on $m$. } \label{fig:5}
\end{figure}

\begin{figure}[h]
\centering{
\includegraphics[width=8cm]{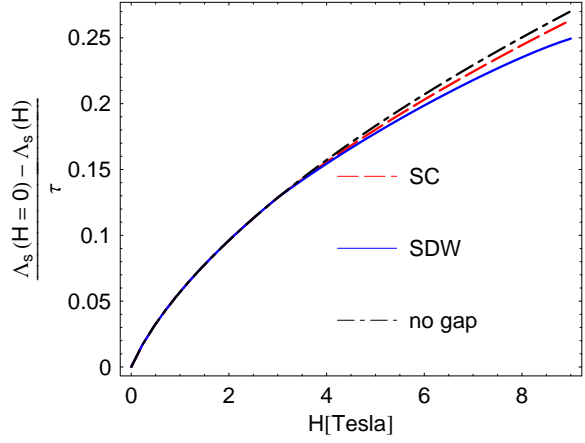}}
\caption{(Color online) The  dependence
$(\Lambda_s(0)-\Lambda_s(H))/\tau$ at $T=0$ for $d$-wave
superconductor with additional SDW, $id_{xy}$ gaps (see
Eq.~(\ref{gap-my}) and Fig.~\ref{fig:5}) and also without any
competing order.} \label{fig:6}
\end{figure}

Substituting into (\ref{J-SDW-H}) $m(H)$ and $\Gamma(H)$ shown in
Fig.~\ref{fig:5} for the SDW case and into (\ref{J0-SC-H})
$\Delta_{d_{xy}}(H)= m(H)$ and $\Gamma = \mbox{const} =
\Gamma(\Delta_{d_{xy}}(H=0))$ for the $i d_{xy}$ case and using
Eq.~(\ref{rho_n-final-H}) after numerical integration over
$\epsilon$ with $\mathcal{P}(\epsilon)$ given by
Eq.~(\ref{P-disorder}) we obtain the results shown in
Fig.~\ref{fig:6}. We use $v_{F}/v_{\Delta}=30$ and $\tau$ as above.
Above $H = 3 \mbox{T}$ the difference between $\rho_s(H)$ without
competing order and $\rho_s(H)$ for the $i d_{xy}$ order is hardly
noticeable, because we used a small value of $b=0.17$. Nevertheless,
for the SDW order $\rho_s(H)$ in high-fields deviates from a
$\sqrt{H}$ behavior quite significantly. This effect is caused by
the decrease of $\Gamma(H)$ seen in Fig.~\ref{fig:5}. We stress that
a similar behavior of $\Lambda_s(H)-\Lambda_s(H=0)$ with a crossover
from $\sqrt{H}$ behavior in low fields to $\ln H$ dependence in high
fields is observed in one of the samples with the lowest $T_c$
studied in Ref.~\cite{Martinoli.report}.

In contrast to Ref.~\cite{Gusynin2004EPJB}, where the decrease of
the thermal conductivity $\kappa(H)$ is directly associated with the
opening of the gap and, accordingly, requires rather large values of
$b$, the results presented here for $\rho_s(H)$ are related to the
indirect influence of the development of the gap on $\Gamma(m(H))$
which in turn leads to the deviations of the dependence $\rho_n(H)$
from a simple $\sim \sqrt{H}$ law.

\section{Conclusions}
\label{sec:concl}

We have addressed the problem of the superfluid density of a
$d$-wave superconductor with competing order. Both the case of a
spin density wave (SDW) and a second (minority) superconducting
order with $i d_{xy}$ symmetry are treated and compared. The nodal
approximation is introduced to treat the main $d$-wave gap and so
the formulation is restricted to low temperatures. The resulting
action which corresponds to QED$_{2+1}$ involves a reducible $4
\times 4$ representation of Dirac  matrices with competing order
equivalent to the different types of Dirac masses. Residual impurity
scattering is accounted for within a $T$-matrix formalism which
includes as special cases the Born and unitary limit.

SDW and a second superconducting order are, in principle, very
different. For example, in the limit when the main $d$-wave order is
set zero, the superfluid density vanishes for the SDW case, but
remains finite for the $i d_{xy}$ (or $i s$) superconducting case.
Also it is found that the SDW order reduces the superfluid density
at $T=0$  (see Eq.~(\ref{Pi-clean-SDW-nodal})) because it competes
for fermi surface with the $d$-wave order, but for $i d_{xy}$ order
there is no such effect. At low temperatures $T \ll m,
\Delta_{d_{xy}}$ the superfluid density acquires an exponentially
activated form (cp. Eqs.~(\ref{rho_n-SDW-exp}) and
(\ref{rho_n-SC-exp})). In the opposite limit $T \ll m,
\Delta_{d_{xy}} \ll \Delta$, where $\Delta $ is the amplitude of the
main $d$-wave gap, the well-known linear in $T$ law
(\ref{rho_n-d-wave}) is modified (see Eqs.~(\ref{rho_n-SDW-asymp})
and (\ref{rho_n-SC-asymp})) and an additional $1/T$ dependence
appears with a coefficient proportional to the square of the
secondary gap value. The sign of the correction is opposite in the
two cases. So far we described results only for the pure limit.
Analytic results are also obtained for the case when impurity
scattering is in the limit when the zero frequency value of the
impurity self-energy $\Gamma \gg T$. The expressions properly reduce
to the known results when the secondary gap is set to zero. In the
both cases impurities modify the classic linear in $T$ dependence to
a $T^2$ dependence as is also the case in $d$-wave superconductor
without competing order. The coefficient of $T^2$ term (see
Eqs.~(\ref{rho_n-SDW-final}) and (\ref{rho_n-SC-final})) however is
modified by the presence of the secondary gap and this modification
is different for SDW and $i d_{xy}$ superconducting order. This is
also the case for the change of the zero temperature limit of the
superfluid density due to impurity scattering.

We have also considered the influence of the opening of a secondary
gap on the magnetic field dependence of the superfluid density for
$\mathbf{H}$ oriented perpendicular to the CuO$_2$ plane. We have
found that the presence of competing orders causes deviations of the
field dependence of the superfluid density $\rho_s(H)$ from a simple
$\sqrt{H}$ law which is associated with the Doppler shift of the
quasiparticle energy. A new and rather important conclusion is that
this effect is caused not only by the developing competing order,
but also by the influence of this order on the residual impurity
scattering rate at zero frequency. On the experimental side the
superfluid density is a directly measurable quantity
\cite{Jeanneret1989APL,Zuev2005PRL,Martinoli.report}, so that the
effects  discussed here may be well within experimental access.
Another advantage of the superfluid density is that it is not blind
with regard to the different kinds of competing orders as is thermal
transport. However, on a theoretical side, in contrast to the
thermal transport, charge transport is renormalized by the vertex
and Fermi liquid corrections \cite{Durst2000PRB} which were not
considered in the present paper. Accordingly the values of $v_F$,
$v_{\Delta}$, $\Gamma$ and $m$ extracted from the measurements of
the superfluid density may be in disagreement with the values for
the same parameters extracted, for example, from the measurements of
thermal conductivity.

\section{Acknowledgments}
We thank D.G.~Hawthorn  and S.Y.~Li for discussing with us their
experimental results. S.G.Sh. gratefully acknowledges V.P.~Gusynin
and V.A.~Miransky for helpful discussions and P.~Martinoli for
stimulating  his interest in the problem. This work was supported
by the Natural Science and Engineering Research Council of Canada
(NSERC) and by the Canadian Institute for Advanced Research
(CIAR).

\begin{thebibliography}{99}
\bibitem{foot1} The normalized superfluid density is related to
$\lambda(T)$ via $\rho_s(T)/\rho_s(0)=\lambda^2(0)/\lambda^2(T)$.
\end{thebibliography}
\end{document}